\documentstyle[aps,multicol,psfig]{revtex}
\newcommand \beq{\begin{eqnarray}}
\newcommand \eeq{\end{eqnarray}}

\newcommand \la{\raisebox{-.5ex}{$\stackrel{<}{\sim}$}}
\newcommand{\av}[1]{\langle{#1}\rangle}
\begin{document}
\title
{Event-by-Event Fluctuations in Ultrarelativistic Heavy-Ion Collisions}
\author{Gordon Baym$^{\ast+}$ and Henning Heiselberg$^{+}$}
\address{$^{\ast}$Department of Physics, University of Illinois at
Urbana-Champaign, 1110 W. Green St., Urbana, Illinois 61801, USA\\
$^{+}$NORDITA, Blegdamsvej 17, DK-2100 Copenhagen \O, Denmark}
\date{Submitted to Physics Letters B, 15 May 1999}

\maketitle

\begin{abstract}
    Motivated by forthcoming experiments at RHIC and LHC, we study
event-by-event fluctuations in ultrarelativistic heavy-ion collisions in
participant nucleon as well as thermal models.  The calculated physical
observables, including multiplicity, kaon to pion ratios, and transverse
momenta agree well with recent NA49 data at the SPS, and indicate that such
studies do not yet reveal the presence of new physics.  Finally, we present a
simple model of how a first order phase transition can be signaled by very
large fluctuations.
\end{abstract}

\vspace{1cm}



\begin{multicols}{2}

    Central ultrarelativistic collisions at RHIC and LHC are expected to
produce at least $\sim 10^4$ particles, and thus present one with the
remarkable opportunity to analyze, on an event-by-event basis, fluctuations in
physical observables such particle multiplicities, transverse momenta,
correlations and ratios.  Analysis of single events with large statistics can
reveal very different physics than studying averages over a large statistical
sample of events.  The use of Hanbury Brown--Twiss correlations to extract the
system geometry is a familiar application of event-by-event fluctuations in
nuclear collisions~\cite{HBT}, and elsewhere, e.g, in
sonoluminesence~\cite{Trentalange}.  The power of this tool has been
strikingly illustrated in study of interference between Bose-Einstein
condensates in trapped atomic systems \cite{bec}.  Recently NA49 has presented
a prototypical event-by-event analysis of fluctuations in central Pb+Pb
collisions at 158 GeV per nucleon at the SPS, which produce more than a
thousand particles per event~\cite{NA49}.

    Studying event-by-event fluctuations in ultrarelativistic heavy ion
collisions to extract new physics was proposed in a series of papers by the
authors and co-workers~[5-7], in which the analysis of transverse energy
fluctuations in central collisions~\cite{Aa} was used to extract evidence
within the binary collision picture for color, or cross-section, fluctuations.
More recent theoretical papers have focussed on different aspects of these
fluctuations, such as searching for evidence for
thermalization~\cite{NA49,GazMrow}, and critical fluctuations at the QCD
phase transition~\cite{SRS,SRS99}.

    In order to be able to extract new physics associated with fluctuations,
it is necessary to understand the role of expected statistical fluctuations.
Our aim here is to study the sources of these fluctuations in collisions.  As
we shall see, the current NA49 data can be essentially understood on the basis
of straightforward statistical arguments.  Expected sources of fluctuations
include impact parameter fluctuations, fluctuations in the number of primary
collisions, and the results of such collisions, nuclear
deformations~\cite{Aa}, effects of rescattering of secondaries, and qcd color
fluctuations.  Since fluctuations in collisions are sensitive to the amount of
rescattering of secondaries taking place, we discuss in detail two limiting
cases, the participant or ``wounded nucleon model'' (WNM), in which one
assumes that particle production occurs in the individual participant nucleons
and rescattering of secondaries is ignored, and the thermal limit in which
scatterings bring the system into local thermal equilibrium.  Whether
rescatterings increase relative fluctuations through greater production of
multiplicity, transverse momenta, etc., or decrease fluctuations by involving
a greater number of degrees of freedom, is not immediately obvious.  Indeed
VENUS simulations~\cite{werner} showed that rescattering had negligible
effects on transverse energy fluctuations.  As we shall see, both models give
similar results for multiplicity fluctuations.  In the wounded nucleon model
fluctuations arise mainly from multiplicity fluctuations for each participant
and from impact parameter fluctuations.  Limited acceptance also influences
the observed fluctuations.  We calculate in detail statistical fluctuations in
multiplicity, K/$\pi$ ratios, and transverse momentum.  Finally, we show in a
simple model how first order phase transitions are capable of producing very
significant fluctuations.

    {\bf Multiplicity fluctuations}:  Let us first calculate fluctuations in
the participant model, which appears to describe well physics at SPS energies
\cite{NA49}.  In this picture
\beq
  N=\sum_i^{N_p} n_i,
\label{partmult}
\eeq
where $N_p$ is the number of participants and $n_i$ is the number of
particles produced in the acceptance by participant $i$.  In the absence of
correlations between $N_p$ and $n$, the average multiplicity is
$\av{N}=\av{N_p}\av{n}$.  For example, NA49 measures charged particles in the
rapidity region $4<y<5.5$ and finds $\av{N}\simeq270$ for central Pb+Pb
collisions.  Finite impact parameters $(b\la3.5$~fm) as well as surface
diffuseness reduce the number of participants from the total number of
nucleons $2A$ to $\av{N_p}\simeq 350$ estimated from Glauber theory; thus
$\av{n}\simeq0.77$.  Squaring Eq.~(\ref{partmult}) assuming
$\av{n_in_j}=\av{n_i}\av{n_j}$ for $i\ne j$, we find the multiplicity
fluctuations
\beq
   \omega_N =  \omega_n + \av{n}\omega_{N_p}, \label{oN}
\eeq
where in general we write
\beq
 \sigma(y) = \av{y^2}-\av{y}^2 \equiv \av{y}\, \omega_y
\eeq
for any stochastic variable $y$.

    A major source of multiplicity fluctuations per participant, $\omega_n$,
is the limited acceptance.  While each participant produces $\nu$ charged
particles, only a smaller fraction $f=\av{n}/\av{\nu}$ are accepted.  Without
carrying out a detailed analysis of the acceptance, one can make a simple
statistical estimate assuming that the particles are accepted randomly, in
which case $n$ is binomially distributed with $\sigma(n)=\nu f(1-f)$ for fixed
$\nu$.  Including fluctuations in $\nu$ we obtain, similarly to
Eq.~(\ref{oN}),
\beq
   \omega_n = 1-f + f \omega_\nu  \,. \label{on}
\eeq
In NN collisions at SPS energies, the charged particle multiplicity is
$\sim7.3$ and $\omega_\nu\simeq 1.9$ \cite{GamleOle}; thus
$\av{\nu}\simeq 3.7$ and $f\simeq 0.21$ for the NA49 acceptance.
Consequently, we find from Eq.~(\ref{on}) that 
$\omega_n\simeq 1.2$.

    As a consequence of nuclear correlations, which strongly reduce density
fluctuations in the colliding nuclei, the fluctuations $\omega_{N_p(b)}$ in
$N_p$ are very small for fixed impact parameter $b$ \cite{sigfluct}.  Almost
all nucleons in the nuclear overlap volume collide and participate.  [By
contrast, the fluctuations in the number of binary collisions is
non-negligible.] Cross section fluctuations play a small role in the WNM
\cite{sigfluct}.  Fluctuations in the number of participants can arise when
the target nucleus is deformed, since the orientations of the deformation axes
vary from event to event \cite{NA34}.  The fluctuations, $\omega_{N_p}$, in
the number of participants are dominated by the varying impact parameters
selected by the experiment.  In the NA49 experiment, for example, the zero
degree calorimeter selects the 5\% most central collisions, corresponding to
impact parameters smaller than a centrality cut on impact parameter,
$b_c\simeq 3.5$ fm.  We have
\beq
  \omega_{N_p}\av{N_p} = \frac{1}{\pi b_c^2}
  \int_0^{b_c} d^2b N_p(b)^2 -\av{N_p}^2 \,,
\eeq
where $\av{N_p}=(1/\pi b_c^2)\int_0^{b_c} d^2b N_p(b)$.  The number of
participants for a given centrality, calculated in \cite{v2}, can be
approximated by $N_p(b)\simeq N_p(0)(1-b/2R)$ for $0\le b\la 3.5$~fm; thus
\beq
  \omega_{N_p} = \frac{N_p(0)}{18} \left(\frac{b_c}{2R}\right)^2
  \,.\label{oNp}
\eeq
For NA49 Pb+Pb collisions with $N_p(0)\simeq 400$ and $(b_c/2R)^2\simeq
5\%$ we find $\omega_{N_p}\simeq 1.1$.  Impact parameter fluctuations are thus
important even for the centrality trigger of NA49.  Varying the centrality cut
or $b_c$ to control such impact parameter fluctuations (\ref{oNp}) should
enable one to extract better any more interesting intrinsic fluctuations.
Recent WA98 analyses confirm that fluctuations in photons and pions grow
approximately linearly with the centrality cut \cite{WA98}.  The Gaussian
multiplicity distribution found in central collisions changes for minimum bias
to a plateau-like distribution \cite{Aa}.

    Calculating $\omega_N$ for the NA49 parameters, we find from Eq.
(\ref{oN}), $\omega_N\simeq 1.2+(0.77)(1.1)=2.0$, in good agreement with
experiment, which measures a multiplicity distribution $\propto
\exp[-(N-\av{N})^2/2\av{N}\omega_N^{exp}]$, where $\omega_M^{exp}$ is of order
$2.01$ \cite{NA49}.

    Let us now consider, in the opposite limit of considerable rescattering,
fluctuations in thermal models.  In a gas in equilibrium, the mean number of
particles per bosonic mode $n_a$ is given by
\beq
  \langle n_a\rangle = \left(\exp{(E_a/T)}-1\right)^{-1} \,, \label{f}
\eeq
with fluctuations
\beq
  \omega_{n_a} = 1+\langle n_a \rangle \,.
\eeq
The total fluctuation in the multiplicity, $N=\sum_a n_a$, is
\beq
 \omega^{BE}_N = 1 + \sum_a\langle n_a\rangle^2/\sum_a\langle n_a\rangle .
 \label{oBE}
\eeq
If the modes are taken to be momentum states, the resulting fluctuations
are $\omega^{BE}_N=\zeta(2)/\zeta(3)=1.37$ for massless particles, while for
pions at temperature $T=150$~MeV $\omega^{BE}_N=1.11$ \cite{Bertsch}.

    Resonances add to fluctuations in the thermal limit whereas they are
implicitly included in the WNM fluctuations.  In high energy nuclear
collisions, resonance decays such as $\rho\to 2\pi$, $\omega\to 3\pi$, etc.,
lead to half or more of the pion multiplicity.  Only a small fraction
$r\simeq10$\% produce two {\it charged} particles in a thermal hadron gas
\cite{Koch} or in RQMD \cite{HH}.  Including such resonance fluctuations in
the BE fluctuations gives, similarly to Eq.  (\ref{oN}),
\beq
 \omega^{BE+R}_N = r\frac{1-r}{1+r} + (1+r)\omega_N^{BE}
\eeq
With $r\simeq 0.1$ we obtain $\omega^{BE+R}_N\simeq 1.3$.  If not all of
the decay particles fall into the NA49 acceptance the fluctuations from
resonances will be reduced.  In \cite{SRS99} the estimated effect of
resonances is about twice ours:  $\omega_N\simeq 1.5$, not including impact
parameter fluctuations.

    Fluctuations in the effective collision volume add a further term
$\av{N}\sigma(V)/\av{V}^2$ to $\omega^{BE+R}_N$.  Assuming that the volume
scales with the number of participants,
$\omega_V/\av{V}\simeq\omega_{N_p}/\av{N_p}$, we find from Eq.~(\ref{oN}) that
$\omega_N=\omega^{BE+R}_N+\av{n}\omega_{N_p}\simeq 2.1$, again consistent with
the NA49 data.  Because of the similarity between the magnitudes of the
thermal and WNM multiplicity fluctuations, the present measurements cannot
distinguish between these two limiting pictures.

    {\bf Kaon/pion ratio}:  To second order in the fluctuations of the numbers
of K and $\pi$, we have
\beq
 \av{K/\pi} =\frac{\av{K}}{\av{\pi}}\left(1+\frac{\omega_\pi}{\av{\pi}} -
               \frac{\av{K\pi}- \av{K}\av{\pi}}{\av{K}\av{\pi}}   \right).
\eeq
The corresponding fluctuations in $\av{K/\pi}$ are given by
\beq
  D^2\equiv \frac{\omega_{K/\pi}}{\av{K/\pi}} =
  \frac{\omega_K}{\av{K}}+ \frac{\omega_\pi}{\av{\pi}}
     -2\frac{\av{K\pi}- \av{K}\av{\pi}}{\av{K}\av{\pi}} \,.
     \label{D}
\eeq
The fluctuations in the kaon to pion ratio is dominated by the
fluctuations in the number of kaons alone.  The third term in Eq.~(\ref{D})
includes correlations between the number of pions and kaons.  It contains a
negative part from volume fluctuations, which removes the
volume fluctuations in $\omega_K$ and $\omega_\pi$ since such fluctuations
cancel in any ratio.  In the NA49 data
\cite{NA49} the average ratio of charged kaons to charged pions is
$\av{K/\pi}=0.18$ and $\av{\pi}\simeq 220$.  Excluding volume fluctuations, we
take $\omega_K\simeq\omega_\pi\simeq 1.2-1.3$ as discussed above.  The first
two terms in Eq.~(\ref{D}) then yield $D\simeq 0.20$ in good agreement with
preliminary measurements $D=0.23$ \cite{NA49}.  Thus at this stage the data
gives no evidence for correlated production of K and $\pi$, as described by
the final term in Eq.~(\ref{D}), besides volume fluctuations.  The similar
fluctuations in mixed event analyses $D_{mixed}=0.208$ \cite{NA49} confirm
this conclusion.

    {\bf Transverse momentum fluctuations}:  The total transverse momentum per
event
\beq
P_t=\sum_{i=1}^N p_{t,i} \,,
\eeq
is very similar to the transverse energy, for which fluctuations have been
studied extensively~\cite{Aa,PRL91}.  The mean transverse momentum and
inverse slopes of distributions generally increase with centrality or
multiplicity.  Assuming that $\alpha\equiv d\log(\av{p_{t}}_N)/d\log N$ is
small, as is the case for pions \cite{NA44slopes}, the average transverse
momentum per particle for given multiplicity $N$ is to leading order
\beq
   \av{p_{t}}_N = \av{p_t}(1+\alpha (N-\av{N})/\av{N}) \,.
\eeq
where $\av{p_t}$ is the average over all events of the single particle
transverse momentum.  With this parametrization, the average total transverse
momentum per particle in an event obeys $\av{P_t/N}=\av{p_t}$.  When the
transverse momentum is approximately exponentially distributed with inverse
slope $T$ in a given event, $\av{p_{t,i}}=2T$, and
$\sigma(p_{t,i})=2T^2=\av{p_t}^2/2$.  This latter fluctuation is in principle
dependent on multiplicity, but as a higher order effect, we ignore it in the
following.

    The total transverse momentum per particle in an event has fluctuations
\begin{eqnarray}
 \av{N}\sigma(P_t/N) &=& \sigma(p_{t}) + \alpha^2 \av{p_t}^2 \omega_N
    \nonumber \\
  &+&  \av{\frac{1}{N}  \sum_{i\ne j}(p_{t,i}p_{t,j} -\av{p_{t}}^2)}
  \,.  \label{opt}
\end{eqnarray}
The three terms on the right are respectively:

    {\rm i)} The individual fluctuations $\sigma(p_{t,i})=\av{p_{t,i}^2} -
\av{p_{t}}^2$, the main term.  In the NA49 data, $\av{p_t}=377$~MeV and
$\av{N}=270$.  From Eq.~(\ref{opt}) we thus obtain
$(\sigma(P_t/N)^{1/2}/\av{p_t}\simeq1/\sqrt{2\av{N}}=4.3$\%, which accounts
for most of the experimentally measured fluctuation 4.65\% \cite{NA49}.  The
data contains no indication of intrinsic temperature fluctuations in the
collisions.

    {\rm ii)} Effects of correlations between $p_t$ and $N$, which are
suppressed with respect to the first term by a factor $\sim \alpha^2$.  In
NA49 the multiplicity of charged particles is mainly that of pions for which
$T\simeq \av{p_{t}}/2$ increases little compared with pp collisions, and
$\alpha\simeq 0.05-0.1$.  Thus, these correlations are small for the NA49 data.
However, for kaons and protons, $\alpha$ can be an order of magnitude larger
as their distributions are strongly affected by the flow observed in central
collisions \cite{NA44slopes}.

    {\rm iii)} Correlations between transverse momenta of different particles
in the same event.  In the WNM the momenta of particles originating from the
same participant are correlated.  In Lund string fragmentation, for example, a
quark-antiquark pair is produced with the same $p_t$ but in opposite
direction.  The average number of pairs of hadrons from the same participant
is $\av{n(n-1)}$ and therefore the latter term in Eq.~(\ref{opt}) becomes
$(\av{n(n-1)}\av{n}) (\av{p_{t,i}p_{t,j\ne i}}-\av{p_{t}}^2)$. To a good
approximation, $n$ is Poisson distributed, i.e., $\av{n(n-1)}/\av{n}=\av{n}$,
equal to 0.77 for the NA49 acceptance, so that this latter term becomes
$\simeq (\av{p_{t,i}p_{t,j\ne i}}-\av{p_{t}}^2)$.  The momentum correlation
between two particles from the same participant is expected to be a small
fraction of $\sigma(p_{t,i})$.

    To quantify the effect of rescatterings, Ref.~\cite{GazMrow} suggested
studying the differences in $\av{N}\sigma(P_t/N)$ and $\sigma(p_{t})$ via the
quantity
\beq
 \Phi(p_t) \simeq \sqrt{\av{N}\sigma(P_t/N)} - \sqrt{\sigma(p_{t,i})}
  \label{pij} \,.
\eeq
As we see from Eq.~(\ref{opt}), in the applicable limit that the second
and third terms are small,
\beq
    \Phi(p_t) &\simeq& \frac{1}{\sqrt{\sigma(p_{t,i})}}
    \left(\alpha^2 \av{p_t}^2 \omega_N +
             (\av{p_{t,i}p_{t,j\ne i}}-\av{p_{t}}^2)\right)  \,.\nonumber\\
  && \label{phii}
\eeq
In the Fritiof model, based on the WNM with no rescatterings between
secondaries, one finds $\Phi(p_t)\simeq 4.5$~MeV.  In the thermal
limit the correlations in Eq.~(\ref{pij}) should vanish for classical
particles but the interference of identical particles (HBT
correlations) contribute to these correlations by $\sim 6.5$~MeV
\cite{Mrow} and slightly reduced by resonances. 
The NA49 experimental value, $\Phi(p_t)=5$~MeV
(corrected for two-track resolution) seems to favor the thermal limit
\cite{NA49}.  Note however that with $\alpha\simeq 0.05-0.1$, the
second term on the right side of Eq.~(\ref{phii}) alone leads to $\Phi
\simeq 1-4$~MeV, i.e., the same order of magnitude.  If
$(\av{p_{t,i}p_{t,j\ne i}}-\av{p_{t}}^2)$ is not positive, then one
cannot a priori rule out that the smallness of $\Phi(p_t)$ does not
arise from a cancellation of this term with $\alpha^2 \av{p_t}^2
\omega_N$, rather than from thermalization.

    {\bf First order phase transitions} can lead to rather large fluctuations
in physical quantities. Thus, detection of enhanced fluctuations, beyond
the elementary statistical ones considered to this point, could signal the
presence of such a transition.  For example, matter undergoing a transition
from chirally symmetric to broken chiral symmetry could, when expanding,
supercool and form droplets, resulting in large multiplicity versus rapidity
fluctuations \cite{HJ}.  Let us imagine that $N_D$ droplets fall into the
acceptance, each producing $n$ particles, i.e., $\av{N}=\av{N_D}\av{n}$.  The
corresponding multiplicity fluctuation can be computed analogously to Eq.
(\ref{oN})
\beq
   \omega_N = \omega_n + \av{n} \omega_{N_D} \,. \label{oND}
\eeq
    As in Eq.~(\ref{on}), we expect $\omega_n\sim 1$.  However, unlike the
case of participant fluctuations, the second term in (\ref{oND}) can lead
to huge multiplicity fluctuations when only a few droplets fall into the
acceptance; in such a case, $\av{n}$ is large and $\omega_{N_D}$ of order
unity.  The fluctuations from droplets depends on the total number of
droplets, the spread in rapidity of particles from a droplet, $\delta y\sim
\sqrt{T/m_t}$, as well as the experimental acceptance in rapidity, $\Delta y$.
When $\delta y\ll \Delta y$ and the droplets are binomially distributed in
rapidity, $\omega_{N_D}\simeq 1-\Delta y/y_{\rm tot}$, which can be a
significant fraction of unity.

    In the extreme case where none or only one droplet falls into the
acceptance with equal probability, we have $\omega_{N_D}=1/2$ and
$\av{n}=2\av{N}$.  The resulting fluctuation is $\omega_N\simeq\av{N}$, which
is {\it more than two orders of magnitude larger} than the expected value of
order unity as currently measured in NA49.  This simple example clearly
demonstrates the importance of event-by-event fluctuations accompanying phase
transitions, and illustrates how monitoring such fluctuations versus
centrality becomes a promising signal, in the upcoming RHIC experiments, for
the onset of a transition.

    This work was initiated at the 13th Nordic Meeting on Intermediate and
High Energy Nuclear Physics, Graeftaavallen, Sweden, and supported in part by
National Science Foundation Grants No.  PHY94-21309 and PHY98-00978.  We are
grateful to A.D.  Jackson for discussions, and Volker Koch and Sergei Voloshin
for helpful comments.


\end{multicols}

\end{document}